\documentclass[12pt]{article}
\usepackage{a4wide,epsfig,amsmath,amssymb,cite,scalefnt,graphicx}
\def\prl{Phys.\ Rev.\ Lett.\ }

\def\epsilonbar{\overline{\epsilon}}

\def\be{\begin{equation}}
\def\ee{\end{equation}}
\def\besub{\begin{subequations}}
\def\eesub{\end{subequations}}
\def\bea{\begin{eqnarray}}
\def\eea{\end{eqnarray}}

\def\qbar{\overline q}

\def\pa{\partial}

\def\Ncal{{\cal N}}

\def\Kcal{{\cal K}}

\def\Fbar{\overline F}
\def\ibar{\overline i}
\def\jbar{\overline j}
\def\kbar{\overline k}
\def\lbar{\overline l}
\def\mbar{\overline m}
\def\nbar{\overline n}
\def\pbar{\overline p}
\def\qbar{\overline q}

\def\Ocal{{\cal O}}
\def\Otcal{\tilde\Ocal}

\def\hbar{{\overline h}}

\def\mbar{\overline{m}}

\def\Phibar{{\overline \Phi}}
\def\thetabar{{\overline\theta}}
\def\alphadot{\dot\alpha}

\def\Qbar{\overline Q}
\def\psib{\overline{\psi}}

\def\varphib{\overline{\varphi}}
\def\Phib{\overline\Phi}

\newcommand{\beq}{\begin{equation}}
\newcommand{\eeq}{\end{equation}}
\newcommand{\beqa}{\begin{eqnarray}}
\newcommand{\eeqa}{\end{eqnarray}}

\def\tr{{\rm tr}}
\def\lsl{\not{\hbox{\kern-1.7pt $\ell$}}}
\def\ksl{\not{\hbox{\kern-2.1pt $k$}}}
\def\Psl{\not{\hbox{\kern-2.1pt $P$}}}

\def\be{\begin{equation}}
\def\bea{\begin{eqnarray}}
\def\eea{\end{eqnarray}}
\def\nn{\nonumber\\}

\begin{document}

\begin{titlepage}
\begin{flushright}
LTH784\\ 
\end{flushright}

\vspace*{3mm}

\begin{center}
{\Huge
One-loop divergences in the two-dimensional non-anticommutative 
supersymmetric $\sigma$-model}\\[12mm]
{\bf I.~Jack and R. Purdy}\\

\vspace{5mm}
Dept. of Mathematical Sciences,
University of Liverpool, Liverpool L69 3BX, UK\\

\end{center}

\vspace{3mm}
\begin{abstract}
We discuss the structure of the non-anticommutative $\Ncal=2$ 
non-linear $\sigma$-model in two dimensions, constructing differential 
operators which implement the deformed supersymmetry generators 
and using them to reproduce the classical action. We then compute the
one-loop quantum corrections and express them in a more compact form using
the differential operators.  
\end{abstract}

\vfill

\end{titlepage}

\section{Introduction}
The subject of deformed quantum field theories has attracted 
renewed attention in recent years due to their natural appearance 
in string theory. Initial studies were devoted to theories on 
non-commutative spacetime in which the commutator of the spacetime 
co-ordinates becomes
non-zero. More recently\cite{casal}-\cite{oog}, 
non-anticommutative supersymmetric theories have been 
constructed by deforming the anticommutators of the grassman co-ordinates
$\theta^{\alpha}$
(while leaving the anticommutators of the $\thetabar{}^{\alphadot}$ unaltered).
Consequently, the anticommutators of the supersymmetry generators 
$\Qbar_{\alphadot}$ are
deformed while those of the $Q_{\alpha}$ are unchanged. 
Non-anticommutative versions of the Wess-Zumino model and supersymmetric gauge
theories have been formulated in four 
dimensions\cite{seiberg,araki} and their renormalisability
discussed\cite{brittoa}-\cite{lunin}, with explicit computations up to two 
loops\cite{grisa} for the Wess-Zumino 
model and one loop for gauge theories\cite{jjwa}-\cite{jjwc}. 

More recently 
still, 
non-anticommutative theories in two dimensions have been considered. On the 
one hand non-anticommutative versions of particular non-linear $\sigma$-models 
have been constructed (by dimensional reduction from four 
dimensions)\cite{inami} and the one-loop corrections computed\cite{arakib}; 
on the 
other hand a non-anticommutative version of the general $\Ncal=2$ 
K\"ahler $\sigma$-model has been constructed directly in two dimensions, 
initially in Refs.~\cite{chand, chanda} but then given an elegant
reformulation in Refs.~\cite{luis,chandc}. We shall predominantly 
follow the notation of Ref.~\cite{luis}, where
the deformation was interpreted as a ``smearing'' of the K\"ahler potential. 
The undeformed $\Ncal=2$ K\"ahler $\sigma$-model and its renormalisation were
studied exhaustively in the context of string theory.  It was thought for a 
while that its only divergences were at the one-loop level where they can be 
interpreted as a correction to the K\"ahler metric of the form of the Ricci 
tensor; until explicit calculations\cite{grisc,grisd} revealed a 
divergence at the 
four-loop level. 

The motivation for the present work was to investigate 
whether the one-loop corrections in the deformed theory as presented in
Ref.~\cite{luis} would exhibit a similar
``smearing'' as in the classical theory. It turns out that the number of 
one-loop diagrams in the deformed theory is enormous, at least in the 
component formulation in which we work; however, they can be expressed in
terms of differential operators implementing the undeformed supersymmetry
generators $Q_{\pm}$ (using light-cone co-ordinates in two dimensions), 
acting on a simpler ``kernel''. Now in fact, the undeformed 
classical action (in its component form) can be expressed simply as the 
product of 
the operators representing {\it all} the supersymmetry generators, $Q_{\pm}$ 
and $\Qbar_{\pm}$,  acting on the K\"ahler potential.
This inspired the hope that in the non-anticommutative case,
if we could construct the operators implementing 
the deformed supersymmetry generators $\Qbar_{\pm}$, we might be able to 
obtain a similarly 
succinct form for the deformed one-loop corrections. Accordingly, we start
by giving an exact construction for these operators to all orders in the 
deformation parameter. We then give our results for the one-loop calculation, 
expressed in a relatively compact form in terms of the undeformed operators 
for $Q_{\pm}$ 
acting on a kernel $\Kcal$. It is then easy to see that unfortunately it is 
impossible to further write $\Kcal$ in a shorter form using the operators 
representing $\Qbar_{\pm}$. 

\section{$\Ncal=2$ supersymmetry in two dimensions}
In this section we set the scene for the analysis by describing in some detail
the case of undeformed supersymmetry in two dimensions, focussing on the
use of differential operators to implement the supersymmetry and simplify the
description.
In two dimensions it is convenient to use ``lightcone'' co-ordinates 
$x^{\pm}$, $\theta^{\pm}$,
$\thetabar{}^{\pm}$ (a slight abuse of terminology since in the 
non-anticommutative case we are obliged to work on a spacetime of
Euclidean signature). We now consider a theory with a multiplet of
chiral superfields 
$\Phi^i(x^{\pm},
\theta^{\pm},\thetabar{}^{\pm})$ (with components $\varphi^i, \psi^i, F^i$).
We denote the conjugate fields by $\Phibar^{\ibar}$, $\varphib^{\ibar}$, etc; 
though often we suppress the superscripts. 
The simplest model is the two-dimensional 
$\Ncal=2$ non-linear $\sigma$-model whose action 
is, in (undeformed) superspace, given by
\be
S_0=\int d^2x d^2\theta d^2\thetabar K(\Phi,\Phibar)
\label{eq:claslag}
\ee
where $K$ is the K\"ahler potential. 

The charges are then
\be
Q_{\pm}=\frac{\pa}{\pa\theta^{\pm}},\quad \Qbar_{\pm}=
-\frac{\pa}{\pa\thetabar{}^{\pm}}-i\theta^{\pm}\frac{\pa}{\pa y^{\pm}},
\ee
where
\be
y^{\pm}=x^{\pm}-i\theta^{\pm}\thetabar{}^{\pm}.
\label{eq:ydef}
\ee
They satisfy the algebra
\bea
Q_+^2=Q_-^2&=&\{Q_+,Q_-\}=0,\nn
\Qbar_+^2=\Qbar_-^2=0,&\quad&\{\Qbar_+,\Qbar_-\}=0,\nn
\{\Qbar_+,Q_+\}=-i\pa_+, &\quad& \{\Qbar_-,Q_-\}=-i\pa_-.\label{eq:undefcom}
\eea
The superfields have expansions in terms of component fields given by
\bea
\Phi&=&\varphi+\theta^+\psi_++\theta^-\psi_-+\theta^+\theta^-F,\nn
\Phibar&=&\varphib+\thetabar{}^+\left[\psib_+-i\theta^+\pa_+\varphib\right]
+\thetabar{}^-\left[\psib_--i\theta^-\pa_-\varphib\right]\nn
&+&\thetabar{}^+\thetabar{}^-\left[\Fbar
+i\theta^+\pa_+\psib_--i\theta^-\pa_-\psib_+
+\theta^+\theta^-\pa_+\pa_-\varphib\right],
\label{compexp}
\eea
where the component fields are functions of $y^{\pm}$, as defined in 
Eq.~(\ref{eq:ydef}).
It is useful to represent the charges $Q_{\pm}$, $\Qbar_{\pm}$
by differential operators $q_{\pm}$, $\qbar_{\pm}^0$ acting
on the fields, i.e.
\besub
\bea
\left[Q_{\pm},\Phi\right]&=&q_{\pm}\Phi,\label{eq:undefopaa}\\
\left[\Qbar_{\pm},\Phi\right]&=&\qbar{}^0_{\pm}\Phi
\label{eq:undefopab}
\eea
\eesub
where
\bea
q_{\pm}&=&\psi_{\pm}\frac{\pa}{\pa\varphi}\mp F\frac{\pa}{\pa\psi_{\mp}}
-i\pa_{\pm}\varphib\frac{\pa}{\pa\psib_{\pm}}\pm i\pa_{\pm}\psib_{\mp}
\frac{\pa}{\pa\Fbar},\nn
\qbar{}^0_{\pm}&=&-\psib_{\pm}\frac{\pa}{\pa\varphib}\pm \Fbar\frac{\pa}
{\pa\psib_{\mp}}
+i\pa_{\pm}\varphi\frac{\pa}{\pa\psi_{\pm}}\mp i\pa_{\pm}\psi_{\mp}
\frac{\pa}{\pa F}.
\label{eq:undefopb}
\eea
We use the superscript ``$0$" to denote the undeformed case; since $q_{\pm}$
will be unchanged in the deformed case, no superscript is
needed for the unbarred operators. 
These operators have anticommutation properties analogous to 
Eq.~(\ref{eq:undefcom}), except that 
\be
\{\qbar^0_+,q_+\}=i\pa_+,\quad \{\qbar^0_-,q_-\}=i\pa_-.
\ee
Note the change in sign; the origin of this can be seen by commuting 
Eqs.~(\ref{eq:undefopaa}), (\ref{eq:undefopab}) with $\Qbar_{\pm}$,
$Q_{\pm}$ respectively and using 
\be
[q_{\pm},\Qbar_{\pm}]=[\qbar^0_{\pm},Q_{\pm}]=0
\ee
(which follows from 
\be
[q_{\pm},\pa_{\pm}]=[\qbar^0_{\pm},\pa_{\pm}]=0)
\ee
in conjunction with 
\be
[A,[B,C]]+[B,[A,C]]=[\{A,B\},C]
\ee
and Eq.~(\ref{eq:undefcom}).
 
The transformations of $\Phi$, $\Phib$ induced by 
$\epsilon^+Q_++\epsilon^-Q_-+\epsilonbar^+\Qbar_++\epsilonbar^-
\Qbar_-$ are then given by
\bea
\delta\Phi&=&[\epsilon^+Q_++\epsilon^-Q_-+\epsilonbar{}^+\Qbar_++
\epsilonbar{}^-\Qbar_-,\Phi],\\
\delta\Phibar&=&[\epsilon^+Q_++\epsilon^-Q_-+\epsilonbar{}^+\Qbar_++
\epsilonbar{}^-\Qbar_-,\Phibar],
\eea
which, in view of Eq.~(\ref{compexp}), entails
\bea
\delta\varphi&=&\epsilon^+\psi_++\epsilon^-\psi_-,\nn
\delta\psi_+&=&\epsilon^-F+i\epsilonbar^+\pa_+\varphi,\nn
\delta\psi_-&=&-\epsilon^+F+i\epsilonbar^-\pa_-\varphi,\nn
\delta F&=&-i\epsilonbar^+\pa_+\psi_-+i\epsilonbar^-\pa_-\psi_+\nn
\delta\varphib&=&-\epsilonbar^+\psib_+-\epsilonbar^-\psib_-,\nn
\delta\psib_+&=&-i\epsilon^+\pa_+\varphib-\epsilonbar^-\Fbar,\nn
\delta\psib_-&=&-i\epsilon^-\pa_-\varphib+\epsilonbar^+\Fbar,\nn
\delta\Fbar&=&i\epsilon^+\pa_+\psib_--i\epsilon^-\pa_-\psib_+.
\eea
By virtue of Eqs.~(\ref{compexp}),
(\ref{eq:undefopaa}), (\ref{eq:undefopab}) we can also
write 
\be
\delta\varphi=(\epsilon^+q_++\epsilon^-q_-+\epsilonbar{}^+\qbar_++
\epsilonbar{}^-\qbar_-)\varphi,
\label{varp}
\ee
with similar expressions for the other component fields.

The effect of the $\int d^2\theta d^2\thetabar$ in Eq.~(\ref{eq:claslag}) is 
to yield the 
component action as the $\theta^2\thetabar{}^2$ term in the expansion of
$K(\Phi,\Phibar)$, giving
\bea
S_0&=&\int d^2x\Bigl[
K_{\jbar}\pa_+\pa_-\varphib{}^{\jbar}+K_{\jbar\kbar}\pa_
+\varphib{}^{\jbar}\pa_-\varphib{}^{\kbar}\nn
&+&
K_{i\jbar}\Bigl(i\psi_+^i\pa_-\psib{}_+^{\jbar}
+i\psi^i_-\pa_+\psib{}^{\jbar}_-+F^i\Fbar{}^{\jbar}\Bigl)\nn
&-&K_{ik\jbar}\psi_+^i\psi_-^k\Fbar{}^{\jbar}
-K_{\ibar\kbar j}\psib{}_+^{\ibar}\psib{}_-^{\kbar}
F^j
+iK_{i\jbar\kbar}\Bigl(\psi_+^i\psib{}^{\jbar}_
+\pa_-\varphib{}^{\kbar}+\psi_-^i\psib{}_-^{\jbar}\pa_+\varphib{}^{\kbar}\Bigr)\nn
&+&K_{ij\ibar\jbar}\psi^i_+\psi_-^j
\psib{}_+^{\ibar}\psib{}_-^{\jbar}\Bigr],
\label{undefact}
\eea
where $K_i=\frac{\pa K}{\pa \varphi^i}$. It is
easily verified using Eqs.~(\ref{eq:undefopb}),
(\ref{undefact}), that 
\be
q_{\pm}S_0=\qbar^{0}_{\pm}S_0=0,
\label{actinv}
\ee
which demonstrates the invariance of the action under supersymmetry 
transformations (according to Eq.~(\ref{varp})).
 
The action Eq.~(\ref{undefact}) can also be written using the operators 
$q_{\pm}$, $\qbar{}^0_{\pm}$ as 
\be
S_0=\int d^2xq_-q_+\qbar{}^0_-\qbar{}^0_+K;
\label{eq:clasker}
\ee
which of course guarantees Eq.~(\ref{actinv}) due to
the nilpotency of $q_{\pm}$, $\qbar^0_{\pm}$, which in turn 
follows from that of $Q_{\pm}$, $\Qbar_{\pm}$ in Eq.~(\ref{eq:undefcom}). 
There is
something intriguingly reminiscent of the BRST formalism in the use of 
nilpotent operators to obtain an invariant expression.
It is worth mentioning that after eliminating the auxiliary fields $F$,
$\Fbar$ using their equations of motion, the action may be written in the
form
\bea
S_0&=&\int d^2x\Bigl[g_{i\jbar}\left(\pa_
+\varphi^{i}\pa_-\varphib{}^{\jbar}+i\psi_+^i\pa_-\psib{}_+^{\jbar}
+i\psi^i_-\pa_+\psib{}^{\jbar}_-\right)\nn
&+&R_{i\ibar j\jbar}i\psi^i_+\psi_-^j\psib{}_+^{\ibar}\psib{}_-^{\jbar}\Bigr]
\label{eq:covform}
\eea
where $R_{i\ibar j\jbar}$ is the Riemann tensor constructed from the
K\"ahler metric $g_{i\jbar}\equiv K_{i\jbar}$. This form is manifestly 
generally covariant with respect to this metric.

At the quantum level the renormalisation of the model may be achieved by
replacing the classical K\"ahler potential by a bare version, $K_B$, chosen 
so as to cancel the ultraviolet divergences order by order. Using the
standard dimensional regularisation with the spacetime dimension continued 
to $d=2-\epsilon$, at one loop we have simply
\be
K_B=K+\frac{1}{2\pi\epsilon}\tr\ln K_{i\jbar}.
\ee
This corresponds to replacing the K\"ahler metric by 
\be
g_{B i\jbar}=g_{i\jbar}+\frac{1}{2\pi\epsilon} R_{i\jbar}
\ee
where $R_{i\jbar}$ is the Ricci tensor. 
As mentioned in the introduction, the next divergence appears at the four-loop
level\cite{grisc,grisd}. 
Just as the classical action may be obtained by the operators 
$q_{\pm}$, $\qbar{}^0_{\pm}$ acting on $K$ as in Eq.~(\ref{eq:clasker}),
we may write 
\be
S_{0B}=\int d^2xq_-q_+\qbar{}^0_-\qbar{}^0_+K_B,
\label{eq:quanker}
\ee
so that in particular $q_-q_+\qbar{}^0_-\qbar{}^0_+\tr\ln K_{i\jbar}$
has the effect of reproducing the one-loop divergences, in a somewhat
compact form.

\section{Non-anticommutative supersymmetry in two dimensions}
In this section we repeat the analysis of the previous section for the case
of deformed two-dimensional supersymmetry.
For the deformed version we take
\be
(\theta^{\pm})^2=(\thetabar{}^{\pm})^2=0,\quad
\{\thetabar{}^+,\thetabar{}^-\}=0,\quad \{\theta^+,\theta^-\}=\frac1M.
\ee
The charges then satisfy the algebra
\bea
Q_+^2=Q_-^2&=&\{Q_+,Q_-\}=0,\nn
\Qbar_+^2=\Qbar_-^2=0,&\quad&\{\Qbar_+,\Qbar_-\}
=-\frac4M\frac{\pa^2}{\pa y^+\pa y^-},\nn
\{\Qbar_+,Q_+\}=-i\pa_+, &\quad& \{\Qbar_-,Q_-\}=-i\pa_-.\label{eq:acomm}
\eea
The non-anticommutativity is implemented at the level of 
superfields by introducing the star-product, which satisfies
\bea
\theta^+*\theta^-=\theta^+\theta^-+\frac{1}{2M},&\quad&
\theta^-*\theta^+=-\theta^+\theta^-+\frac{1}{2M},\nn
\theta^+*\theta^+\theta^-=-\frac{1}{2M}\theta^+,&\quad&
\theta^-*\theta^+\theta^-=\frac{1}{2M}\theta^-,\nn
\theta^+\theta^-*\theta^+\theta^-&=&\frac{1}{4M^2}.
\eea
We now wish to construct differential operators $\qbar_{\pm}$ representing 
the effects of $\Qbar_{\pm}$ in the deformed case in a similar manner to
Eq.~(\ref{eq:undefopaa},\ref{eq:undefopab}),
extending $\qbar_{\pm}^0$ given in Eq.~(\ref{eq:undefopb}) for the undeformed 
case. (The operators $q_{\pm}$ are unchanged by the deformation.)
We start by examining the effects of $\Qbar_{\pm}$ on powers of 
$\Phi$ alone, since mixed products of $\Phi$ and $\Phibar$ present 
additional complications. Defining
\be
I_r^{(n)}=\int_{-\frac12}^{\frac12}d\xi\left(\frac{\xi}{M}\right)^r
\left(\varphi+\frac{\xi}{M}F\right)^n
\ee
it is straightforward to show using the methods of Ref.~\cite{luis} that
\be
\Phi^n_*=
(1+\theta^+q_+)(1+\theta^-q_-)\left(I_0^{(n)}-q_+q_-I_1^{(n)}\right),
\ee
where $\Phi^n_*$ denotes the star-product of $n$ $\Phi$'s.
Then acting on $\Phi^n_*$, $\Qbar_{\pm}$ are represented by
\bea
\qbar^{\Phi}_+&=&\qbar{}^{0}_+-\frac{i}{2M}\pa_{+}q_-+i(
-q'_+q'_-[\pa'_+q'_-]\Otcal+\pa'_+q'_-\Ocal+
[\pa'_+q'_-]\Ocal),\nn
\qbar^{\Phi}_-&=&\qbar{}^{0}_--\frac{i}{2M}\pa_{-}q_+
-i(-q'_+q'_-[\pa'_-q'_+]\Otcal+\pa'_-q'_+\Ocal+
[\pa'_-q'_+]\Ocal). \label{eq:fullq}
\eea
Here a prime denotes the part of the operator containing derivatives
with respect to the chiral (but not the anti-chiral) fields, and
correspondingly
\be
\pa_{\pm}'=\pa_{\pm}\varphi\frac{\pa}{\pa\varphi}+
\pa_{\pm}\psi_+\frac{\pa}{\pa\psi_+}+\pa_{\pm}\psi_-\frac{\pa}{\pa\psi_-}
+\pa_{\pm}F\frac{\pa}{\pa F}.
\ee
Moreover, 
\be
[\pa'_+q'_-]=\pa_+\psi_{-}\frac{\pa}{\pa\varphi}+\pa_+F\frac{\pa}{\pa\psi_{+}},
\ee
and 
\bea
\Ocal I_0^{(n)}&=&I_1^{(n)},\nn
\Ocal I_1^{(n)}&=& I_2^{(n)}-\Otcal I_0^{(n)}.\label{eq:orels}
\eea
These properties are guaranteed by the following definitions: 
\bea
\Ocal&=&\sum_{r=1}^{\infty}a_r
\left(\frac{1}{M^2}\right)^r
\left(F\frac{\pa}{\pa \varphi}\right)^{2r-1},\nn
\Otcal&=&\sum_{r=1}^{\infty}(2r-1)a_r
\left(\frac{1}{M^2}\right)^r
\left(F\frac{\pa}{\pa \varphi}\right)^{2r-2},\eea
where the $a_r$ must satisfy for each $n\ge1$ 
\be
\sum_{r=0}^{n-1}\frac{a_{n-r}}{2^{2r}(2r+1)(2r)!}=\frac{1}{
2^{2n}(2n+1)(2n-1)!}.
\ee
 We have been unable to find a closed form for the $a_r$;
the first few, determined recursively, being
\be
a_1=\frac{1}{12}, \quad a_2=-\frac{1}{720},\quad a_3=\frac{1}{2^5.3^3.5.7}.
\ee
To check that the operators in Eq.~(\ref{eq:fullq}) do indeed
represent the operators $\Qbar_{\pm}$ according to
\be
[\Qbar_{\pm},\Phi^n_*]_*=\qbar^{\Phi}_{\pm}\Phi^n_*
\ee
(where $[\phantom{q},\phantom{q}]_*$ represents the commutator evaluated using
star-products)
we need to use Eqs.~(\ref{eq:orels}) in
conjunction with
\bea
\qbar_+^{0\prime}I_r^{(n)}=-i[\pa'_+q'_-]I_{r+1}^{(n)},&\quad&
 \qbar_-^{0\prime}I_r^{(n)}=i[\pa'_-q'_+]I_{r+1}^{(n)},\nn
q''_+I_0^{(n)}=q''_-I_0^{(n)}&=&q''_+I_1^{(n)}=q''_-I_1^{(n)}=0.\label{eq:idsa}
\eea
where a double prime denotes the part of the operator containing derivatives
with respect to the anti-chiral (but not the chiral) fields.

It is easy to check that the operators in Eq.~(\ref{eq:fullq})
satisfy the anticommutation relations of Eq.~(\ref{eq:acomm}),
using 
\be
\left[\qbar^{0}_{\pm},F\frac{\pa}{\pa \varphi}\right]=\mp i[\pa'_{\pm}q'_{\mp}]
\ee
(which implies
\be
[\qbar_{\pm}^0,\Ocal]=\mp i[\pa'_{\pm}q'_{\mp}]\Otcal).
\ee 

When acting on products of both $\Phi$ and $\Phibar$ the situation is more 
complicated, and the operators representing $\Qbar_{\pm}$ will require
modification. We have $\Phibar{}_*^n=\Phibar{}^n$ and we find
\bea
\Phi_*^n*\Phibar{}^m&=&
(1+\theta^+q_+)(1+\theta^-q_-)\nn
&&\left[1-\thetabar{}^+\left(\qbar^{0\prime\prime}_+
-\frac{i}{2M}\pa_+''q_-'\right)\right]
\left[1-\thetabar{}^-\left(\qbar^{0\prime\prime}_-
-\frac{i}{2M}\pa_-''q_+'\right)\right]\nn
&&\left(I_0^{(n)}-q_+q_-I_1^{(n)}\right)\varphib{}^m.\label{eq:unsymma}
\eea
We then have
\bea
\left[\Qbar_+,\Phi^n_**\Phibar{}^m\right]_*
&=&\left\{\qbar^{\Phi}_+-\frac{i}{2M}(\pa''_+q'_--\pa'_+q''_-)\right\}\Phi^n_**
\Phibar{}^m,\nn
\left[\Qbar_-,\Phi^n_**\Phibar{}^m\right]_*
&=&\left\{\qbar^{\Phi}_--\frac{i}{2M}(\pa_-''q_+'-\pa_-'q_+'')\right\}\Phi^n_**
\Phibar{}^m.\label{eq:opuna}
\eea
On the other hand we have
\bea
\Phibar{}^m*\Phi_*^n&=&
(1+\theta^+q_+)(1+\theta^-q_-)\nn
&&\left[1-\thetabar{}^+\left(\qbar''_+
+\frac{i}{2M}\pa_+''q_-'\right)\right]
\left[1-\thetabar{}^-\left(\qbar''_-
+\frac{i}{2M}\pa_-''q_+'\right)\right]\nn
&&\left(I_0^{(n)}-q_+q_-I_1^{(n)}\right)\varphib{}^m,\label{eq:unsymmb}
\eea
and correspondingly
\bea
\left[\Qbar_+,\Phibar{}^m*\Phi^n_*\right]_*
&=&\left\{\qbar^{\Phi}_++\frac{i}{2M}(\pa''_+q'_--\pa'_+q''_-)\right\}
\Phibar{}^m*\Phi^n_*,\nn
\left[\Qbar_-,\Phibar{}^m*\Phi^n_*\right]_*
&=&\left\{\qbar^{\Phi}_-+\frac{i}{2M}(\pa_-''q_+'-\pa_-'q_+'')\right\}
\Phibar{}^m*\Phi^n_*.\label{eq:opunb}
\eea
We see from Eqs.~(\ref{eq:opuna}), (\ref{eq:opunb}) that the 
operators representing $\Qbar_{\pm}$ are modified in different ways depending 
on whether they act on $\Phi^n_**\Phibar{}^m$ or $\Phibar{}^m*\Phi^n_*$. 
It is unusual to find that the representation of the operator depends on 
the ordering of the term on which it acts. However, fortunately 
we are only interested
in the deformed version of the K\"ahler potential, in which each term should be
defined as a symmetrised star-product of $\Phi$'s and $\Phib$'s, and therefore
the ordering question will not arise. For such
a symmetrised product, the representations of $\Qbar_{\pm}$ will again be 
different from those given in Eq.~(\ref{eq:fullq}).  
For an undeformed K\"ahler potential  
\be
K[\Phi,\Phibar]=\sum_{n,m}K_{n,m}\Phi^n\Phibar{}^m,
\ee
the natural definition of the deformed K\"ahler potential is 
\be
K_*[\Phi,\Phibar]=\sum_{n,m}K_{n,m}[\Phi^n\Phibar{}^m]_*,
\ee
where $[\Phi^n\Phibar{}^m]_*$ represents the symmetrised star-product of 
$n$ $\Phi$'s and $m$ $\Phibar$'s. It can be shown that
\bea
K_*[\Phi,\Phibar]&=&
(1+\theta^+q_+)(1+\theta^-q_-)\left(1-\thetabar{}^
+\qbar^{0\prime\prime}_+\right)
\left(1-\thetabar{}^-\qbar^{0\prime\prime}_-\right)\nn
&&\left[K_0(\varphi,F,\varphib)-q_+q_-K_1(\varphi,F,\varphib)\right]\nn
&-&\frac{1}{4M^2}\thetabar{}^+\thetabar{}^-q'_+q'_-\pa''_+\pa''_-
K_0(\varphi,F,\varphib),
\label{Kstar}
\eea
where
\be
K_m(\varphi,F,\varphib)=\int_{-\frac12}^{\frac12}d\xi\xi^mK\left(\varphi+
\frac{\xi}{M}F,\varphib\right).
\ee
The symmetrisation has resulted in the disappearance of most of the terms 
involving a $\frac{1}{2M}$ in Eqs.~(\ref{eq:unsymma}), (\ref{eq:unsymmb}).
Correspondingly we no longer need the $\frac{1}{2M}$ terms in 
Eqs.~(\ref{eq:opuna}), (\ref{eq:opunb}). 
However, the residual $\frac{1}{4M^2}$ term requires a modification of the 
operators given in Eqs.~(\ref{eq:fullq}), so that
\bea
\qbar_+&=&\qbar{}^{0}_+-\frac{i}{2M}\pa_{+}q_--\frac{i}{4M^2}
(\pa_+''q_+'q_-'q_-''+\pa_+'q_-'q_+''q_-'')\nn
&+&i(-q'_+q'_-[\pa'_+q'_-]\Otcal+\pa'_+q'_-\Ocal+
[\pa'_+q'_-]\Ocal),\nn
\qbar_-&=&\qbar{}^{0}_--\frac{i}{2M}\pa_{-}q_++\frac{i}{4M^2}
(-\pa_-''q_+'q_-'q_+''+\pa_-'q_+'q_+''q_-'')\nn
&-&i(-q'_+q'_-[\pa'_-q'_+]\Otcal+\pa'_-q'_+\Ocal+
[\pa'_-q'_+]\Ocal), \label{eq:fullqa}
\eea
We can verify that these 
operators do indeed implement the operators $\Qbar_{\pm}$ according to 
\be
[\Qbar_{\pm},K_*]_*=\qbar_{\pm}K_*,
\ee
using
the analogue of Eq.~(\ref{eq:orels}) for the K\"ahler potential, 
\bea
\Ocal K_0&=&K_1,\nn
\Ocal K_1&=& K_2-\Otcal K_0\label{eq:orelsa}
\eea
together with the analogue of Eq.~(\ref{eq:idsa}),
\bea
\qbar_+'K_r=-i[\pa'_+q'_-]K_{r+1},&\quad&
\qbar_-'K_r=i[\pa'_-q'_+]K_{r+1},\nn
q''_+K_0=q''_-K_0&=&q''_+K_1=q''_-K_1=0.
\eea

The action is given by the $\theta^2\thetabar{}^2$ term and hence from 
Eq.~(\ref{Kstar})
\be
S=\int d^2x q_-q_+\qbar_-''\qbar_+''(K_0-q_+q_-K_1),
\label{defactb}
\ee
which can be expanded as\cite{chand} -\cite{luis}
\bea
S&=&\int d^2x
\Bigl\{\pa_{\jbar}{K_0}\pa_+\pa_-\varphib{}^{\jbar}+\pa_{\jbar}\pa_{\kbar}{K_0}\pa_
+\varphib{}^{\jbar}\pa_-\varphib{}^{\kbar}
+
\pa_i\pa_{\jbar}{K_0}\Bigl(i\psi_+^i\pa_-\psib{}_+^{\jbar}
+i\psi^i_-\pa_+\psib{}^{\jbar}_-+F^i\Fbar{}^{\jbar}\Bigl)\nn
&-&\pa_i\pa_k\pa_{\jbar}{K_0}\psi_+^i\psi_-^k\Fbar{}^{\jbar}
-\pa_{\ibar}\pa_{\kbar}\pa_j{K_0}\psib{}_+^{\ibar}\psib{}_-^{\kbar}
F^j
+i\pa_i\pa_{\jbar}\pa_{\kbar}{K_0}\Bigl(\psi_+^i\psib{}^{\jbar}_
+\pa_-\varphib{}^{\kbar}+\psi_-^i\psib{}_-^{\jbar}\pa_+\varphib{}^{\kbar}\Bigr)\nn
&+&\pa_i\pa_j\pa_{\ibar}\pa_{\jbar}{K_0}\psi^i_+\psi_-^j
\psib{}_+^{\ibar}\psib{}_-^{\jbar}
+\frac{1}{M}\Bigl(\pa_i\pa_{\jbar}{K_1}F^i\pa_+\pa_-\varphib{}^{\jbar}
-\pa_i\pa_k\pa_{\jbar}{K_1}\psi_+^i\psi_-^k\pa_+\pa_-\varphib{}^{\jbar}\nn
&+&\pa_i\pa_{\jbar}\pa_{\kbar}{K_1}F^i\pa_+\varphib{}^{\jbar}\pa_-\varphib{}^{\kbar}
-\pa_i\pa_k\pa_{\jbar}\pa_{\kbar}{K_1}\psi_+^i\psi_-^k\pa_+\varphib{}^{\jbar}\pa_
-\varphib{}^{\kbar}\Bigr)\Bigr\}.
\label{eq:defact}
\eea
It can then be checked that also 
\be
S=\int d^2xq_-q_+\qbar_-\qbar_+(K_0-q_+q_-K_1)
=\int d^2xq_-q_+\qbar_-\qbar_+K_0.
\label{defacta}
\ee
Note that in Eq.~(\ref{defactb}), the $K_1$ term is indispensable and is
entirely responsible for the $K_1$ terms in Eq.~(\ref{eq:defact}); while 
in Eq.~(\ref{defacta}), the $K_1$ term is redundant and can be omitted,
leading to a form for the action similar to Eq.~(\ref{eq:clasker}) in the 
undeformed case. The $K_1$ terms in Eq.~(\ref{eq:defact}) are generated 
from Eq.~(\ref{defacta}) by applying Eq.~(\ref{eq:orelsa}).
 
Finally, from Eq.~(\ref{defacta}), we see that (as in the undeformed case)
the nilpotency of $q_{\pm}$,
$\qbar_{\pm}$, which follows from that of $Q_{\pm}$, $\Qbar_{\pm}$ in 
Eq.~(\ref{eq:acomm}), ensure
\be
q_{\pm}S=\qbar_{\pm}S=0;
\label{defactinv}
\ee
so that the deformed action is invariant under the action of $q_{\pm}$ and
$\qbar_{\pm}$. 
\section{One-loop corrections}

Our goal was to investigate the one-loop corrections for the deformed theory,
and see whether they 
could be interpreted in terms of a ``smearing'' of the background geometry 
as at the classical level. It seemed reasonable to do this order by order
in $\frac{1}{M^2}$. (Note that $K_i$ is a power series in $\frac{1}{M^2}$,
starting at $\frac{1}{M^0}$ for $i$ even and $\frac{1}{M}$ for 
$i$ odd). We then had to make a choice of method, 
since 
the computation of the one-loop and higher quantum corrections for the
undeformed K\"ahler $\sigma$-model may be performed in several different ways.
The superspace computation\cite{grisc} is the most efficient, though it has 
the disadvantage that 
it conceals the generally covariant form of the results, i.e. that they can be
expressed in terms of the K\"ahler metric and its associated Riemann tensor
in a generally covariant way. The covariant form of the classical action is 
achieved in the component formulation upon integrating out the auxiliary 
fields, and computations up to four loops have also been carried out in this
formalism\cite{grisd}. Superspace computations in the non-anticommutative 
case have been performed in the four-dimensional 
context\cite{penrom},\cite{grisb} 
but the formalism is technically rather complex; on the other hand, integrating 
out the auxiliary fields in the deformed action Eq.~(\ref{eq:defact}) would be 
difficult and in any case it is no longer clear if general covariance is a 
useful guide. 

Accordingly, we decided to perform the calculation in the  uneliminated
component formulation. However, it rapidly becomes apparent that there
is a plethora of diagrams to consider. We started by computing the
divergences for the set of graphs with a single insertion of a  vertex
with a $\frac{1}{M^2}$ factor derived from a $K_1$ term in
Eq.~(\ref{eq:defact}). We then realised that the divergent contributions
from this set of graphs (numbering about 200) could be expressed much
more concisely as $q_-q_+\Kcal$ for some $\Kcal$ (which we call a kernel). 
With this as a guide,
we were then able to construct the corresponding $\Kcal$ for the full
set of one-loop $\frac{1}{M^2}$ diagrams, explicitly computing only a
small subset of these to serve as a check. Of course, this is
reminiscent of the fact remarked on  earlier that in the undeformed case
the one-loop quantum corrections may be  written in terms of
$q_-q_+\qbar{}^0_-\qbar{}^0_+\tr\ln K_{i\jbar}$. The full kernel,
$\Kcal^{(1)}_B$, is displayed in the Appendix using a convenient
diagrammatic  notation.  It is tempting to wonder if the analogy with
the undeformed case goes  further so that we may write 
\be
S_B^{(1)}=\int d^2x q_-q_+\qbar_-\qbar_+\tilde \Kcal^{(1)}_B
\label{barform} 
\ee 
for some underlying $\tilde \Kcal_B^{(1)}$,  where
$\qbar_{\pm}$ are the deformed  operators constructed in
Eq.~(\ref{eq:fullqa}); indeed this was our motivation  for constructing
these operators in the first instance. Unfortunately this  turns out not
to be the case, as is easily seen: focussing on the set of graphs  in
$\Kcal_B^{(1)} $ with five vertices, four with a single fermion and one with an $F$, it can be seen that the
graphs with six vertices, five with one fermion and one with an $F$,
(and no derivatives) created by
the action of $\qbar_{+}$ on this set  do not cancel. In drawing this
conclusion we can restrict attention to the  effect of $\qbar^0_+$ since
the remaining terms in $\qbar_+$ all contain derivatives.  Since this is
the only source of graphs of this type in $\qbar_+\Kcal^{(1)}_B$, we see that
$\qbar_+\Kcal^{(1)}_B\ne0$ (and by the  same token
$\qbar_-\Kcal^{(1)}_B\ne0$). Therefore $q_-q_+\qbar_+\Kcal^{(1)}_B\ne0$ 
and
$q_-q_+\qbar_-\Kcal^{(1)}_B\ne0$) (consider for instance those graphs
for which $q_-q_+$ simply attaches an $F$ at the vertex already containing 
an $F$); and so $\qbar_+S_B^{(1)}\ne0$,
$\qbar_-S_B^{(1)}\ne0$.
This  immediately implies (due once again
to the nilpotency of $\qbar_{\pm}$) 
that $S_B^{(1)}$ cannot be of the form
Eq.~(\ref{barform}). It is noteworthy that the
classical behaviour is not reproduced at the quantum level, and in
particular that the one-loop effective action is not invariant under
$\qbar_{\pm}$, even though the classical action was.

\section{Conclusions}
We have constructed differential operators which express the 
non-anticommutative supersymmetry according to 
\be
[Q_{\pm},\Phi]=q_{\pm}\Phi,\quad [\Qbar_{\pm},\Phi]=\qbar{}_{\pm}\Phi
\ee
and which therefore reproduce the deformed algebra in Eq.~(\ref{eq:acomm}).
It then follows from the fact that the classical action may be written  
$S=\int d^2x q_-q_+\qbar_-\qbar_+K_0$ and the  
nilpotency of $q_{\pm}$, $\qbar_{\pm}$ that
$q_{\pm}S=\qbar_{\pm}S=0$. However, we then examined the one-loop effective 
action and showed that although we could express the one-loop divergences as
\be
S_B^{(1)}=\int d^2x q_-q_+\Kcal^{(1)}_B,
\ee
it was not possible in turn to write 
$\Kcal^{(1)}_B=\qbar_-\qbar_+\tilde\Kcal^{(1)}_B$ for some
$\tilde\Kcal^{(1)}_B$.
Correspondingly, although $q_{\pm}S^{(1)}_B=0$, $\qbar_{\pm}S^{(1)}_B\ne0$.
In fact, an invariance of the classical action can be shown to lead directly 
to an invariance of the quantum effective action only in simple cases, namely 
for linear transformations of the fields; such as, indeed, the transformations
corresponding to $q_{\pm}$. In the case of non-linear transformations, the 
transformation properties of the effective action are expressed through
Ward identities. In the case at hand, the effect of $\qbar_{\pm}$ on a 
single field is in fact linear, though the effect on functions of the fields
is more complicated. 

  The fact that $q_{\pm}S=0$ implies $q_{\pm}S^{(1)}_B=0$ is therefore easy to 
understand. However, it would also be interesting to try to prove to all
orders the stronger
statement, that $S_B=\int d^2x q_-q_+ \Kcal_B$ for an appropriate $K_B$, which 
we have shown to be valid at one loop and first order in $\frac{1}{M^2}$.
  Our original motivation in embarking on this calculation was to see if the 
``smearing'' of the classical geometry was mirrored at the quantum level. This 
seems unlikely in view of the non-renormalisability of the theory,
manifested here by the appearance of divergent
terms in the one-loop effective action
with, for instance, 6 fermion fields; and in fact one can obtain divergent
diagrams with arbitrary numbers of external legs by inserting chains
of deformed vertices of indefinite length into appropriate
``propagators'' in a given divergent diagram. The $\Ncal=\frac12$ gauge theory
in four dimensions, albeit power-counting non-renormalisable, turned out to
have only a finite number of types of counterterm. This property is associated
with the non-hermiticity of the theory, a generic feature of these        
deformed supersymmetric theories; but in the four-dimensional case this can be
codified as a kind of R-parity\cite{lunin}
 which severely restricts the types of 
counterterm; presumably such an effect is absent in two dimensions.
The combination of non-renormalisability and the novel form of the invariance
seems likely to preclude the possibility of obtaining a succinct form of the 
quantum effective action which could be interpreted in terms of a modification 
of the (smeared) background geometry, though it would be interesting to 
investigate this further. Of course, although we committed ourselves to working
in the component formulation, believing the superspace computation of quantum
corrections to be very unwieldy in the nonanticommutative case, this alternative
might be worth pursuing to see if a simpler form of the results might be 
achieved thereby.
  
\vspace*{1em}

\noindent
{\large\bf Acknowledgements}\\
We thank Tim Jones for helpful comments. RP was supported by STFC through
a graduate studentship.

\appendix

\section{One-Loop Kernel}

We present here in largely diagrammatic form 
the kernel $\Kcal_B^{(1)}$ for the one-loop divergences, 
which are then given by $q_-q_+\Kcal_B^{(1)}$.
Since $\qbar_-\qbar_+K_0=\Fbar^{\ibar}K_{0\ibar}
-\psib_+^{\ibar}\psib_-^{\jbar}K_{0\ibar\jbar}$, with a similar expression
for $K_1$,
the action in Eq.~(\ref{defactb}), and hence the kernel, separates into four 
sections which can separately be written as $q_-q_+$ acting on a kernel.  
The kernel may accordingly be written
\be
\Kcal_B^{(1)}=\frac{\pa^2 L_{M^2}}{\pa F^i\pa\varphib^{\jbar}}K^{i\jbar}
-\frac{\pa^2 L_{M^2}}{\pa F^i\pa F^j}K^{i\kbar}K^{j\lbar}
(K_{\kbar\lbar m}F^m-K_{\kbar\lbar mn}\psi^m\psi^n)+
\frac{1}{24M^2}(A_1+A_2+2A_3+2A_4),
\ee
where
$L_{M^2}$ is the $M^2$ term in the lagrangian of Eq.~(\ref{eq:defact}) 
and $A_{1-4}$ are expressed diagrammatically below, in 
Figs.~\ref{fig3}-\ref{fig8}.

\begin{figure}
\includegraphics{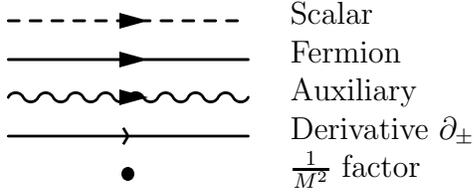}
\caption{Figure conventions}\label{fig1}
\end{figure}

\begin{figure}
\includegraphics{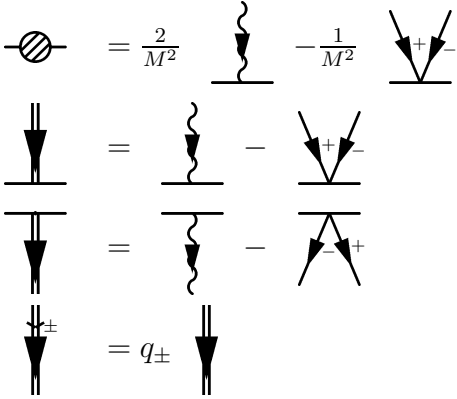}
\caption{Shorthand notation for diagrams}\label{fig2}
\end{figure}

\begin{figure}
\includegraphics{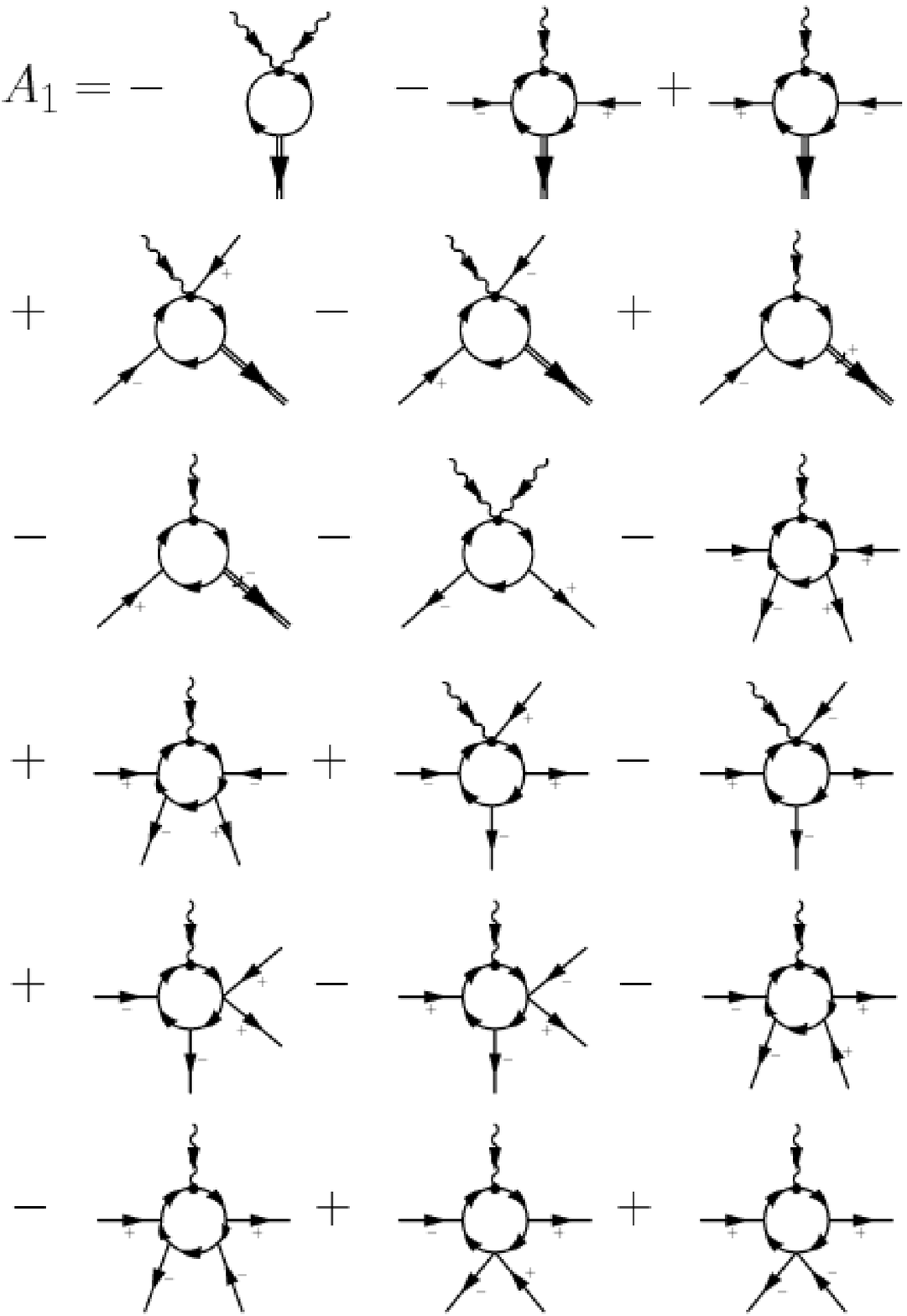}
\caption{Diagrams for $A_1$}\label{fig3}
\end{figure}

In these diagrams a ``propagator'' in a loop denotes $K^{-1}$ and vertices 
denote derivatives of $K$, while external lines attached to vertices 
represent the various fields according to the conventions in Fig.~\ref{fig1}
and the convenient shorthand notations in Fig.~\ref{fig2}.

Incoming (outgoing) arrows represent chiral (antichiral) fields, 
respectively. The ordering of
fermion fields is fixed by the convention that we start at the left-most field
at the top of the diagram and read clockwise around the loop. 
As an illustration
of our notation, the first diagram in $A_1$ in Fig.~\ref{fig3} below represents
\be
F^iF^jK_{ijk\lbar}K^{m\lbar}K^{k\nbar}
(K_{m\nbar\pbar}\Fbar^{\pbar}
-K_{m\nbar\pbar\qbar}\psib_+^{\pbar}\psib_-^{\qbar})
\ee
and the second represents
\be
F^i\psi_+^j\psi_-^kK_{ip\jbar}K^{l\jbar}K_{jl\kbar}K^{m\kbar}
(K_{m\mbar\pbar}\Fbar^{\pbar}
-K_{m\mbar\pbar\qbar}\psib_+^{\pbar}\psib_-^{\qbar})
K^{n\mbar }K_{kn\nbar}K^{p\nbar}
\ee
(where $K^{i\jbar}\equiv K^{-1}_{i\jbar}$).
Using
$\pa_i K^{-1}=-K^{-1}\pa_i KK^{-1}$ the effect of $q_{\pm}$ is to
add external lines and create new vertices. After acting on a diagram with
$q_-q_+$, we obtain a set of diagrams which (unless they cancel with similar 
contributions from other kernel diagrams) correspond to viable one-loop 
Feynman graphs, the vertex with the dot or the ``blob'' being the one from the 
deformed part of the action, and hence with an accompanying $\frac{1}{M^2}$
factor.

\begin{figure}
\includegraphics{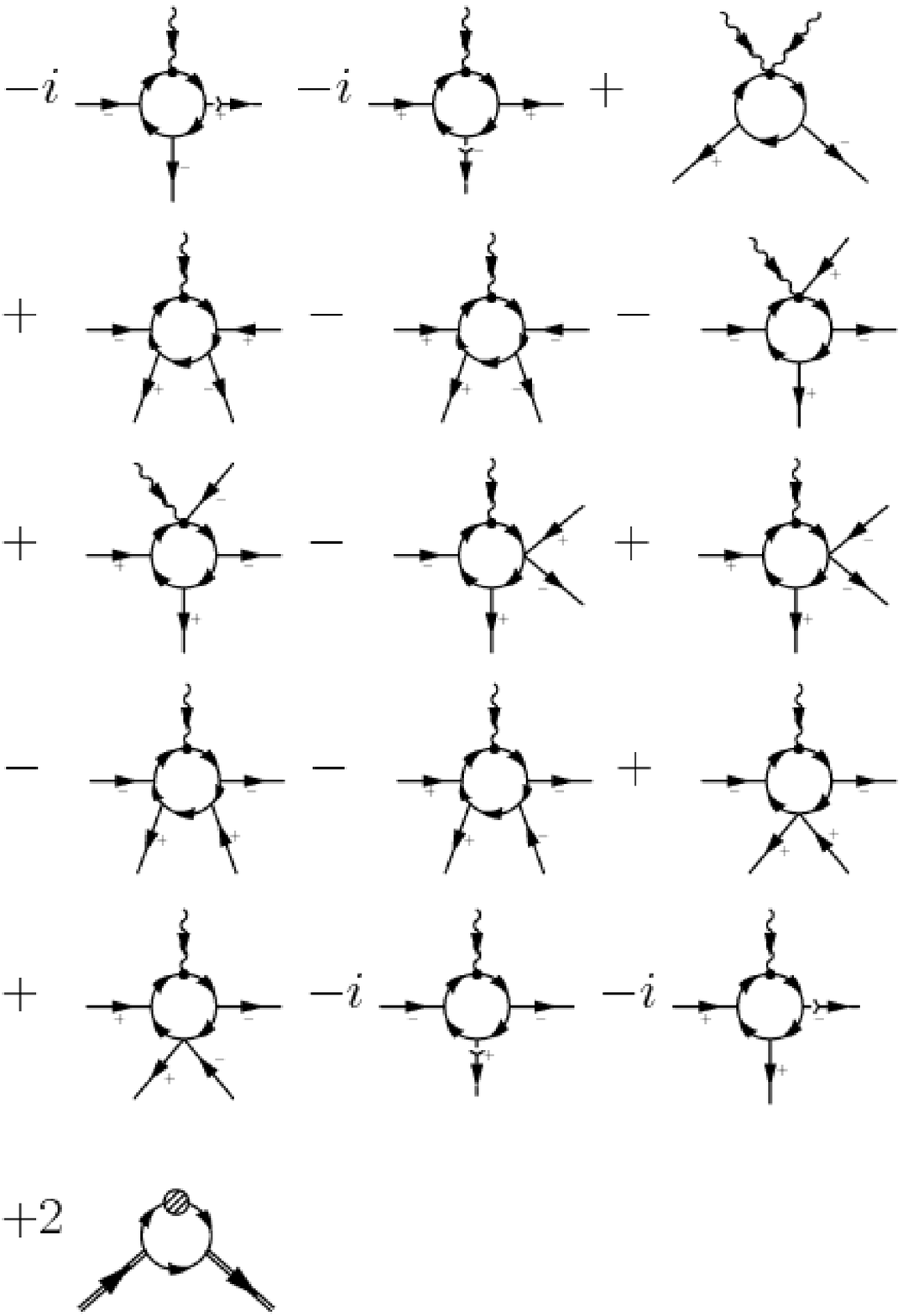}
\caption{Diagrams for $A_1$ (continued)}
\end{figure}

\begin{figure}
\includegraphics{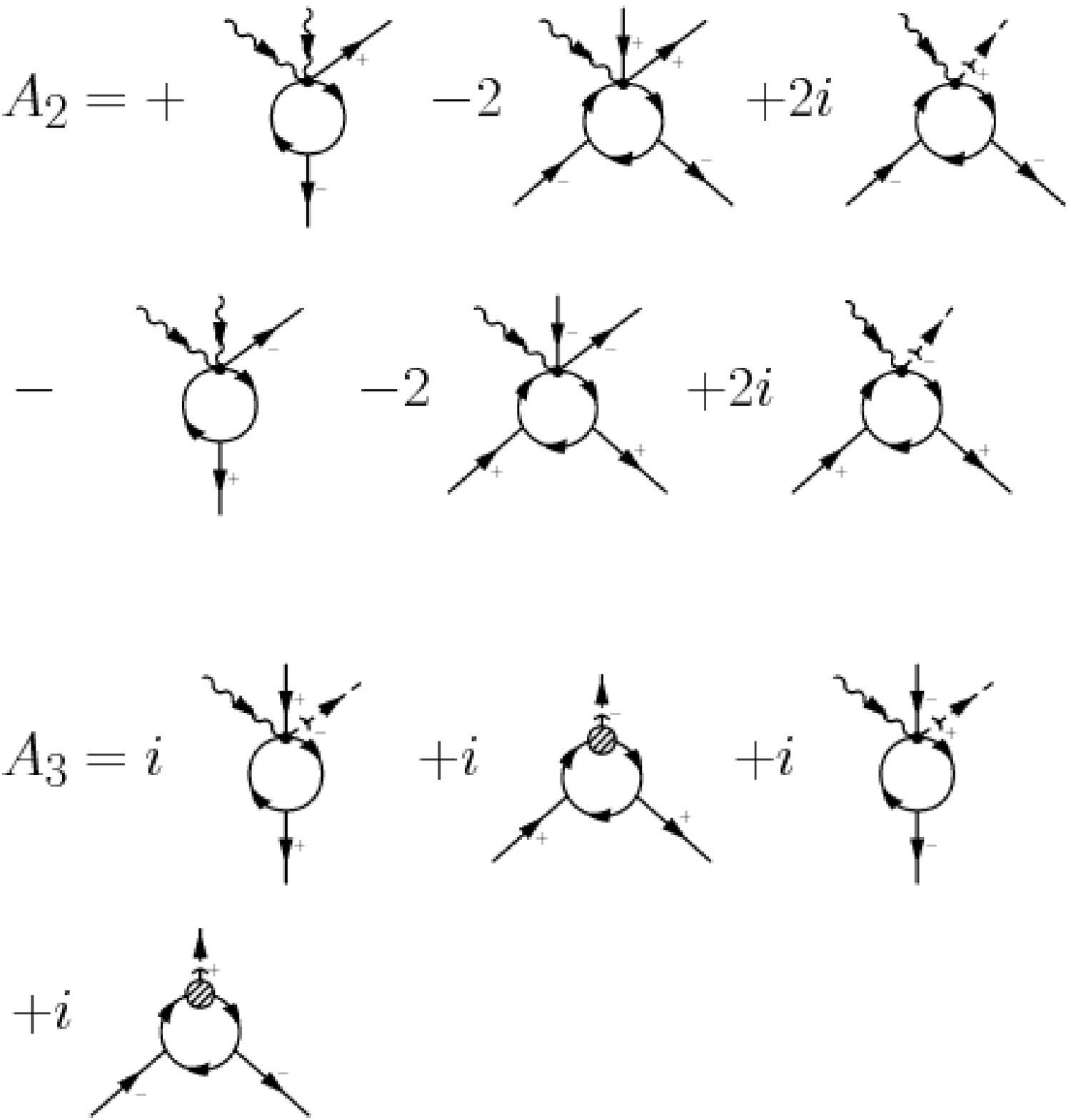}
\caption{Diagrams for $A_2$ and $A_3$}
\end{figure}

\begin{figure}
\includegraphics{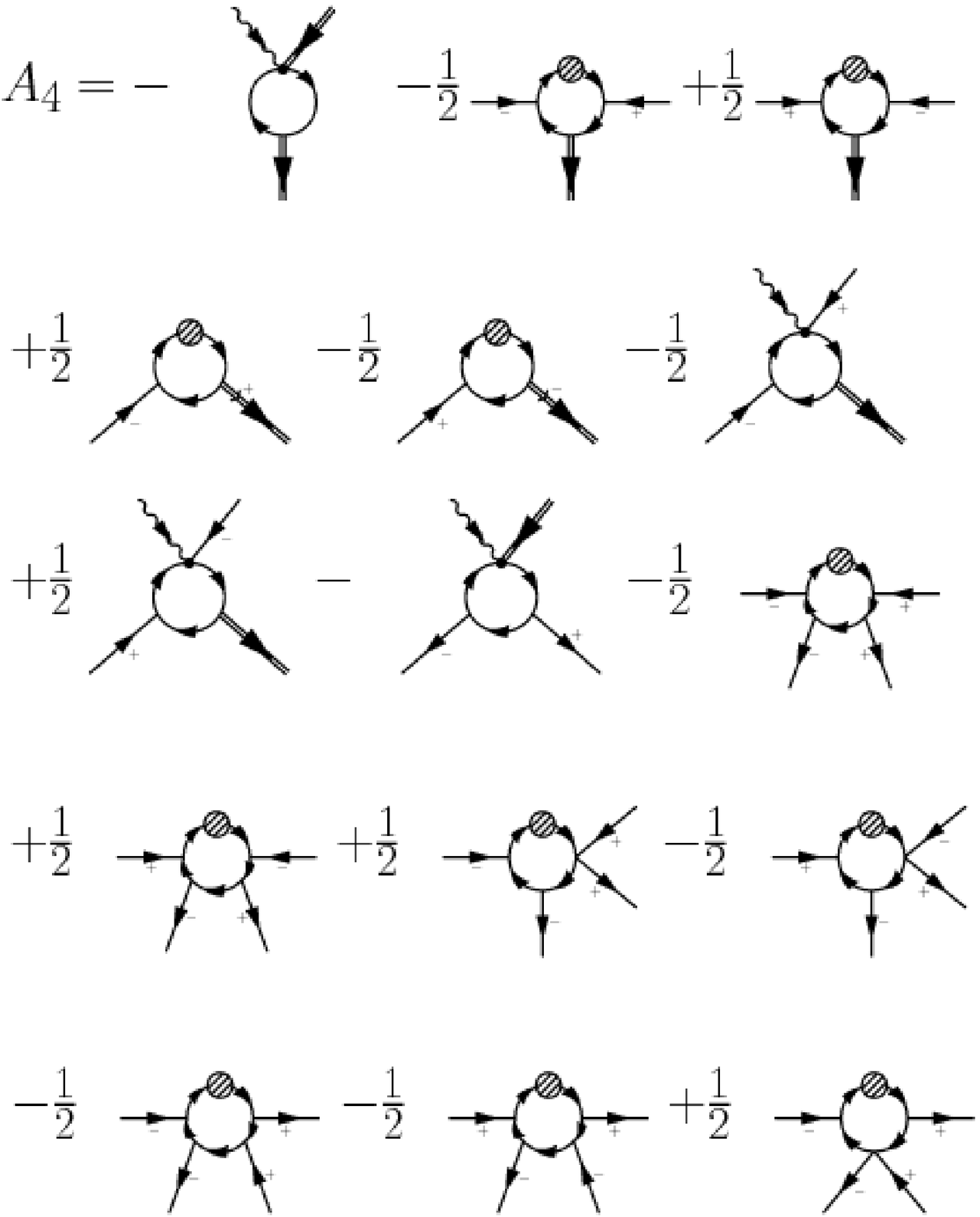}
\caption{Diagrams for $A_4$}
\end{figure}

\begin{figure}
\includegraphics{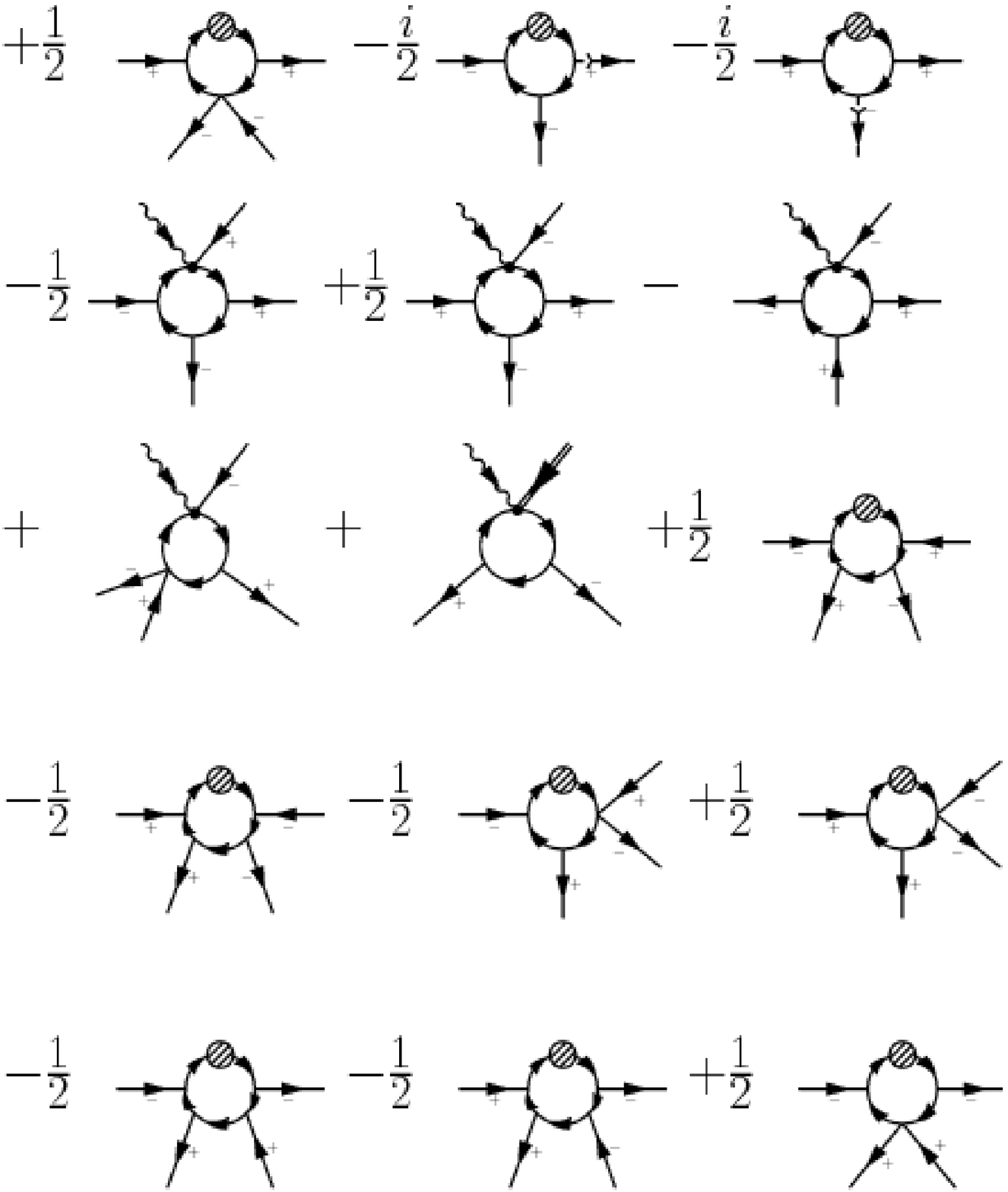}
\caption{Diagrams for $A_4$ (continued)}
\end{figure}

\begin{figure}
\includegraphics{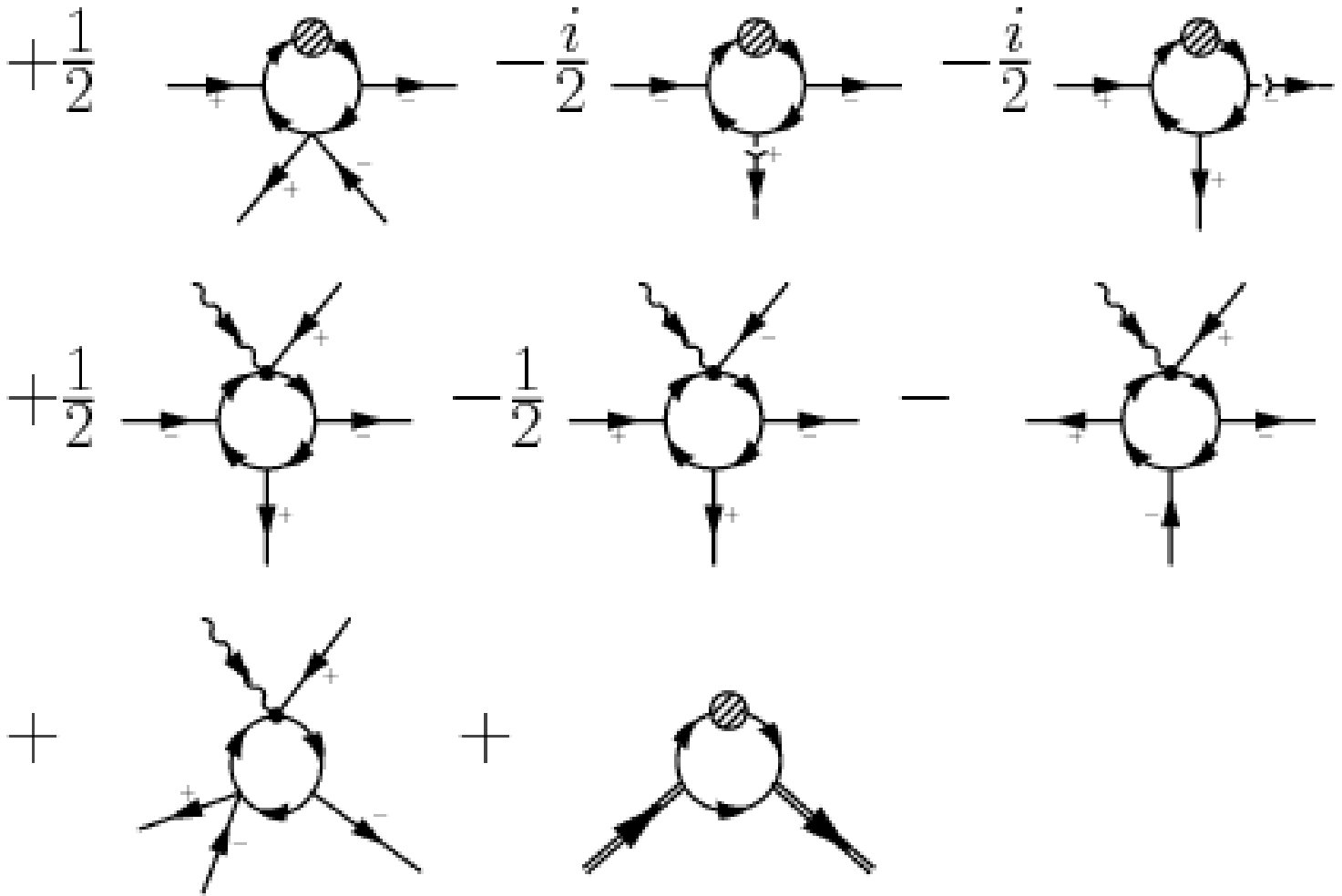}
\caption{Diagrams for $A_4$ (continued)}\label{fig8}
\end{figure}

\bigskip

We observe some intriguing patterns in the groups of diagrams appearing in
$A_{1-4}$ above. For instance, one group of terms in $A_1$ is repeated 
in $A_4$ with the simple substitution of a ``blob'' for an incoming $F$ (and a 
factor of $\frac12$);
and another group of terms in $A_1$ may be obtained from the former group in
$A_1$ by replacing 
a $\psib_+$ followed by an adjacent $\psib_-$ 
(or a $\psib_-$ followed by an adjacent $\psib_+$) with 
a $\Fbar-\psib_+\psib_-$ (i.e. an outgoing double line). Finally, the graphs
in $A_3$ are similar to those of $A_2$.

\end{document}